\documentclass[10pt]{article}
\usepackage[T1]{fontenc}
\usepackage{tabularx}
\usepackage[latin9]{inputenc}
\usepackage{geometry}
\usepackage{array}
\usepackage{verbatim}
\usepackage{float}
\usepackage{rotfloat}
\usepackage{multirow}
\usepackage{amsmath}
\usepackage{amssymb}
\usepackage{hyperref}
\usepackage{graphicx}
\usepackage{booktabs}
\usepackage{esint}
\usepackage{wrapfig}

\usepackage[round,authoryear]{natbib}
\usepackage{amsthm}
\usepackage{amsmath} 
\usepackage{amsfonts}
\usepackage{amssymb}
\usepackage{multibib}
\newcites{own}{Relevant publications by the applicant}
 \DeclareMathOperator*{\E}{\mathbb{E}}
\DeclareMathOperator*{\maxiz}{\mathrm{max}}

\begin{document}


\title{On a computationally-scalable sparse formulation of the multidimensional  and non-stationary maximum entropy principle}
\author
{ Illia Horenko$^{a,\ast}$, Ganna Marchenko$^{a}$, Patrick Gagliardini$^{b,c}$,\\
\small{$^{a}$Universit\`a della Svizzera Italiana, Faculty of Informatics, Via G. Buffi 13, TI-6900 Lugano, Switzerland,}\\
\small{$^{b}$Faculty of Economics, Universit\`a della Svizzera italiana Via G. Buffi 13, CH-6900 Lugano, Switzerland,}\\
\small{$^{c}$Swiss Finance Institute, Walchestrasse 9, CH-8006 Zurich, Switzerland,}\\
\small{$^\ast$To whom correspondence should be addressed; E-mail: horenkoi@usi.ch}\\}

\maketitle
\paragraph{Acknowledgement}
The authors thank Michael Rockinger (HEC Lausanne) for providing the code for PML4 parameter estimator we used for comparison. This work was funded by the Swiss National Research Foundation (Project 140829) and DFG SPP 1114 (Mercator Fellowship of Illia Horenko).
\begin{abstract}
Data-driven modelling and computational predictions based on maximum entropy principle (MaxEnt-principle) aim at finding as-simple-as-possible - but not simpler then necessary - models  that allow to avoid the data overfitting problem. We derive a multivariate non-parametric and non-stationary formulation of the MaxEnt-principle and show that its solution can be approximated through a numerical maximisation of the sparse constrained optimization problem with regularization. Application of the resulting algorithm to popular financial benchmarks reveals memoryless models allowing for simple and qualitative descriptions of the major stock market indexes data. We compare the obtained MaxEnt-models to the  heteroschedastic models from the computational econometrics (GARCH, GARCH-GJR, MS-GARCH, GARCH-PML4) in terms of the model fit, complexity and prediction quality. We compare the resulting model log-likelihoods, the values of the Bayesian Information Criterion, posterior model probabilities, the quality of the data autocorrelation function fits as well as the Value-at-Risk prediction quality. We show that all of the considered seven major financial benchmark time series (DJI, SPX, FTSE, STOXX, SMI, HSI and N225) are better described
by conditionally memoryless MaxEnt-models with nonstationary regime-switching than by the common econometric models with finite memory. This analysis also reveals a sparse network of statistically-significant temporal relations for the positive and negative latent variance changes among different markets. The code is provided for open access. 
\end{abstract}


\baselineskip24pt


Maximum Entropy principle (MaxEnt-principle) was originally introduced in physics and information theory  to search for least-biased probabilistic data descriptions that match certain statistical properties of the data, e.g. the data distribution moments (\cite{Jaynes1957}). The MaxEnt-principle implies that the most unbiased distribution for the data is the one that admits the most uncertainty, measured in terms of entropy. Depending on the constraints imposed in the entropy maximisation problem, different parametric probability distributions that can be described by this principle are Gaussian, Exponential, Laplace, Cauchy, Chi-squared and Gamma distributions,  among others. This MaxEnt-based probabilistic modelling approach has been successfully applied to many problems ranging from biology (\cite{phillips2006maximum}, \cite{mora2010maximum}) and natural language processing (\cite{berger1996maximum}, \cite{nigam1999using}) to applications from economics and finance  (\cite{Zellner88}, \cite{Wu2003}).

In contrast to the parametric MaxEnt-modelling, where the particular parametric distribution models (Gaussian, Exponential, etc.) are dependent on the fixed finite set of constant parameters, computation of nonparametric Maximum Entropy densities even in one dimension is not a trivial task, as was previously discussed in \cite{Agmon79}, \cite{Mead1984} and in \cite{Rockinger_2011}.  In \cite{marchenko2018towards}  a systematic mathematical derivation for a non-stationary extension of the non-parametric MaxEnt-methodology for one-dimensional time series problems was introduced. This BV-Entropy framework imposes a mild bounded-variation (BV) assumption on the time dependence of the non-stationary moments of the underlying non-stationary probability distribution function (p.d.f.). In \cite{marchenko2018towards} it was shown that the original ill-posed non-stationary MaxEnt-principle - formulated as an entropy maximisation problem with time-dependent moment constraints - can be sharply bounded from below via a well-posed and computationally-scalable maximisation problem. This lower-bound problem appeared to be a nonparametric regime-switching entropy maximisation problem with $K$ locally-stationary  regimes, subject to $K$ $l_1$-constraints on the regime-specific vectors of moments and to a BV-constraint for the latent regime transitions. The linear BV-constraint happens to bound from above with $C$ the maximal number of regime switches - and controls the persistence of the obtained MaxEnt regime transition models.   It was also shown that the optimal values of $C$ and $K$ can be estimated deploying common model selection criteria like the Bayesian Information Criterion (BIC). 

However, this model formulation is confined to one-dimensional time series analysis problems only. A direct extension of this one-dimensional BV-Entropy methodology to multiple dimensions is hampered by the curse of dimension and the overfitting problems: linear growth in the number of problem dimension will results in the exponential growth in the number of the underlying mulivariate MaxEnt-parameters that have to be determined from the data statistics. In many practical applications - for example in economics and finance - there is only one historical realisation of the process that is available, with no possibility to obtain other sampling realisations from some "model". For every particular dimension at every particular time there is only one data point available. This means that  a completely nonparametric approach to statistical analysis of such a data leads to ill-posed problems (\cite{Hardle}).

In the following, we describe a computational algorithm achieving a sparse multidimensional extensions of the one dimensional BV-Entropy model from (\cite{marchenko2018towards}), where the coupling between the $n$ individual 1D entropy maximisation problems will be achieved through a sparsifying regularization constraint that controls appropriate function space distances ($l_2$, $l_1$ or BV) between the individual latent factors.

\begin{figure}[h]
\includegraphics[width=1\textwidth]{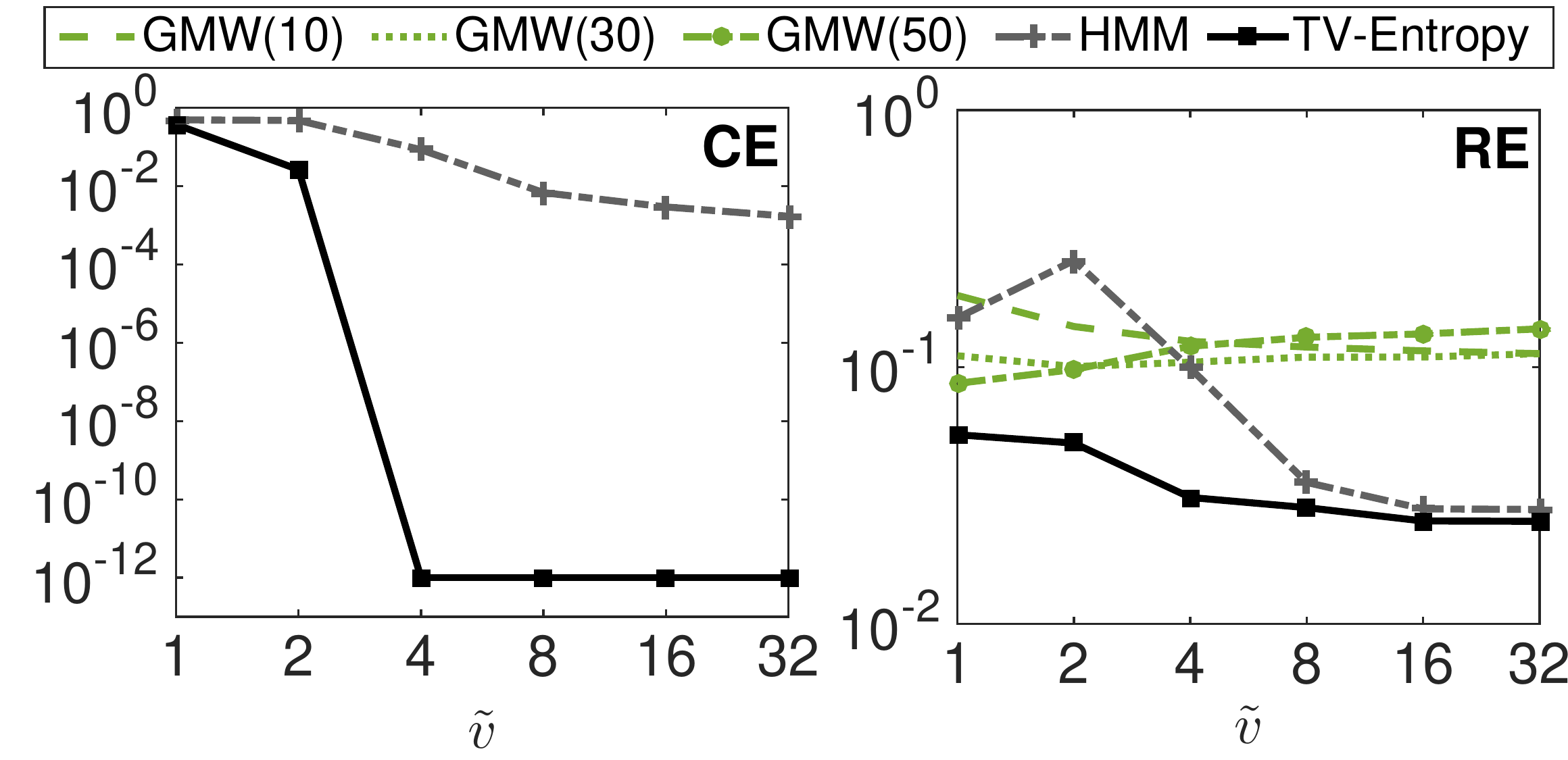}
\caption{\label{errors} Average classification error (CE) between the true and the reconstructed switching regime paths (left) and average relative error (RE) between the true and the reconstructed variance signals (right) in log scale obtained for 100 two-regime Gaussian samples with $\mathcal{N}(0,1)$ and $\mathcal{N}(0,\tilde{v})$ regimes.}
\label{fig:sim_study}
\end{figure}

The resulting algorithm (further referred to as a TV-Entropy) allows a computationally-tractable search for the most unbiased multivariate time-dependent distribution of the data, minimizing the optimal number of the locally-stationary hidden regimes and sparsifying their time-persistent regime-switching dynamic and the regime-specific distribution parameter vectors. Then, we illustrate its performance and compare it to common approaches on the test model system (a regime-switching Gaussian, Figure \ref{fig:sim_study}). Finally, we apply this algorithm and the common GARCH tools to a set of seven popular financial stock market indexes and show that these simpler memoryless maximum entropy models of volatility outperform the popular heteroschedastic methods with finite memory for all of the benchmarks considered (Figures 2-3, Table 2).

\section*{Method}

We start with  $n$-dimensional multivariate time series data $x_{t,i}$, $\forall t \in \{1,\dots,T\}$ and $\forall i \in \{1,\dots,n\}$ on a closed interval $\mathcal{X} = \mathcal{X}_1 \times \cdots \times \mathcal{X}_n  \subset \mathbb{R}^n$. The goal is to find the most descriptive and least complex model for these data. We will first assume that $x_t$ is conditionally independent in time, and would like to estimate its unknown marginal time-dependent probability densities $f_{t,i}(x)$. In the context of the Maximum Entropy principle, the density $f_{t,i}(x)$ can be identified by solving an optimization problem which consists of finding a time-dependent density function with the maximal entropy among all distribution functions that match the data in first $m+1$ sample moments at all of the given time instances $t$ in the dimension $i$. There is a long tradition in statistics and econometrics adopting inference methodologies based on matching sample moments (Generalized Method of Moments (GMM) estimation, \cite{hansen1982large}). Moreover, Maximum Entropy methods are often invoked in multinomial choice problems (\cite{ggolan1996maximum}, \cite{manski1981structural}). Maximisation of the expected entropy with respect to a time-dependent multivariate probability density  $f_{t,i}(x),$  (where the expectation is taken over the time $t$ and the dimension $n$) can be written as:
\begin{align}\label{Hti}
		  \underset{f_{t,i}(x), \, \forall t,i} {\mathrm{max}} \,\, \Bigg\{\E_{t,i} \Big[ H\Big[f_{t,i}(x)\Big] \Big] = \E_{t,i} \Big[  - \int_{\mathcal{X}_i} f_{t,i}(x) \ln f_{t,i}(x) \, dx\Big]\Bigg\},
\end{align} 
subject to
\begin{align}\label{Constr}
	\int_{\mathcal{X}_i}  x^j f_{t,i}(x) \, dx = \mu_{t,i}(j),  \indent	\forall j \in \{0,\dots, m\},\forall i \in \{1,\dots, n\},\forall t \in \{1,\dots, T\}
\end{align} 
and $\mu_{t,i} \in \mathcal{R}^{m+1}$ are time- and dimension-dependent sample moments.
Then, the optimal $f^*_{t,i}(x)$ can be derived by computing first-order optimality conditions of the corresponding optimization problem (1-2), providing the following formulation
\begin{align}\label{MED_t}
	f_{t,i}(x) = \exp \Big[ -\sum_{j=0}^{m} \Lambda_{t,i}(j)x^j \Big], \quad \forall t,i\\
\text{s.t. } 	\int_{\mathcal{X}_i} x^n  \exp \Big[ -\sum_{j=0}^{m} \Lambda_{t,i}(j)x^j \Big]\, dx = \mu_{t,i}(n), \quad \forall n \in \{0, \dots, m\}, \\
 \text{  } \mu_{t,i}(0) = 1, \label{NonStatC}
\end{align} 
where $\Lambda_{t,i} \in \mathcal{R}^{m+1}$ are unknown time-dependent parameters (Lagrange multipliers) of maximum entropy distributions. One way to compute $\Lambda(\cdot)$ would be to maximize the log-likelihood function based on the obtained densities $f_{t,i}(x)$, i.e., 
\begin{align} \label{NonStatLL}
\underset{\Lambda(\cdot)} {\mathrm{max}} \,\, \mathcal{L}\Big(\Lambda(\cdot)\Big) = - \sum_{i=1}^n\sum_{t=1}^T \Big(\sum_{j=1}^{m} \Lambda_{t,i}(j)x_t^j + \ln Z_{\Lambda_{t,i}}\Big),\\
\text{where }
Z_{\Lambda_{t,i}} = \int_{\mathcal{X}_i} \exp \Big[ -\sum_{j=1}^{m} \Lambda_{t,i}(j)x^j \Big] \,dx = \exp[\Lambda_{t,i}(0)],
\end{align}
The first-order optimality conditions of the problem [\ref{NonStatLL}] are equivalent to the conditions [\ref{MED_t}]-[\ref{NonStatC}], therefore $\Lambda(\cdot)$ maximizing the criterion in [\ref{NonStatLL}] are the optimal values of the maximum entropy densities parameters in  [\ref{MED_t}]-[\ref{NonStatC}].

In the realistic applications when only one historical realization sequence $\{x_1,\dots,x_T\}$ is available, there is no straightforward solution to the nonstationary problem [\ref{NonStatLL}], as at each time instance $t$ we have much more parameters than we have observed data. To solve this problem, we are going to introduce two following assumptions.\\
\noindent{\bf Assumption 1:}  \textit{Total variation (a TV norm) of the MaxEnt parameters $\Lambda(\cdot)$ is bounded, i.e.} 
\begin{align} \label{TV-norm}
| \Lambda |_{TV} = \sum_{t_1,t_2=1}^{T} \sum_{i_1,i_2=1}^{n} | \Lambda_{t_1,i_1} (\cdot)- \Lambda_{t_2,i_2} (\cdot)|_1 = C  < + \infty,
\end{align}
\noindent{\bf Assumption 2:} \textit{There exist  $K \ll nT$ distinct sets of parameters $\lambda^{(k)}, k=1,...,K$ and  $\gamma_{t,i} = [\gamma_{t,i}^{(1)}, \dots, \gamma_{t,i}^{(K)}]$ (with $\sum_{k=1}^K\gamma_{t,i}^{(k)}=1$ and $\gamma_{t,i}^{(K)}\geq 0$ for all $t,i,k$), such that for any $t,i$ and $k$  vector $\Lambda_{t,i}$ can be expressed as a convex linear combination} 
\begin{align} \label{Mixture1}
	\Lambda_{t,i} = \sum_{k=1}^K  \gamma_{t,i}^{(k)}\lambda^{(k)}.
\end{align}

Assumptions 1 and 2 introduce sparsity on the $\Lambda_{t,i}$ across time and space indices $t$ and $i$. Very importantly, these assumptions do not rely on any ordering in the space dimension. Indeed, in many practical applications, e.g., in economics and finance  a natural ordering across assets, institutions, markets etc. does not exist. On a practical side, for real financial data these assumptions will be fulfilled automatically if one sets both constants $K$ and $C$ to be large enough (for example, setting $K=nT$ always fulfills Assumption 2). In the practical applications the aim will be in finding the computationally-scalable lower bound estimates of these constants.   
Substituting condition  [\ref{Mixture1}] in [\ref{TV-norm}], getting use of the Jensen's inequality and inserting into [\ref{NonStatLL}]  the obtained inequality constraint as the penalty-term of the Karush-Kuhn-Tucker  leads to the following lower-bound approximation of the MaxEnt-problem   [\ref{NonStatLL}]:  
\begin{eqnarray}
\label{Transformed}
&&\maxiz \limits_{\lambda, \gamma} \indent L(\lambda,\gamma)=-\sum_{k=1}^{\mathbf{K}}  \Bigg[\sum_{i=1;t=1}^{n,T}\gamma_{t,i}^{(k)} \Big (\sum_{j=1}^{m} \lambda^{(k)}_j x_{t,i}^j + Z^{(k)}_i\Big)+\sigma_C  |\lambda^{(k)}|_1\sum_{t_1,t_2=1}^{T} \sum_{i_1,i_2=1}^{n}|\gamma^{(k)}_{t_1,i_1} - \gamma^{(k)}_{t_2,i_2}|\Bigg] \\
\label{c2}
&&\text{s.t. }Z^{(k)}_i=\ln \int_{\mathcal{X}_i} \exp \Bigg[ -\sum_{j=1}^{m} \lambda^{(k)}_{(j)} x^j \Big] \,dx ,\indent\label{c1}  \indent \gamma_{t,i}^{(k)}\geq0, \indent \sum_{k=1}^{\mathbf{K}}\gamma_{t,i}^{(k)} = 1,  \indent \sigma_C\geq 0.
\end{eqnarray}
A solution of the obtained lower-bound problem is an approximation to the solution of the original nonstationary ill-posed Maximum Entropy problem [\ref{Hti}]. The optimization criteria in [\ref{Transformed}] is nonlinear and nonconvex - implying that the problem can have more than one locally-optimal solution. However, there are three properties of this optimization problem formulation that can be exploited numerically: (i) when $\lambda$ is kept fixed,  [\ref{Transformed}]-[\ref{c2}] becomes a uniquely-solvable linear programming (LP) problem w.r.t. $\gamma$  - and can be solved very efficiently by means of common LP-tools (e.g., with the simplex method); (ii) if $\sigma_C=0$, solving [\ref{Transformed}]-[\ref{c2}] becomes equivalent to solving $n$ independent  one-dimensional non-stationary entropy maximisation problems from the BV-Entropy method in \cite{marchenko2018towards};  (iii) algebraic structure of the term containing $\sigma_C$ is similar to the LASSO-regularization formulation very popular in machine learning (\cite{Tibshirani96}), increasing $\sigma_C$ will result in increasing the sparsity of $\lambda$ and in penalizing the temporal and cross-sectional variations in  $\gamma$  - making the  obtained MaxEntropy model approximations more simple, sparse and persistent across dimensions and in time. For a numerical solution of the problem [\ref{Transformed}]-[\ref{c2}] with a fixed set of parameters $K$ and $\sigma_C$ we will adapt the subspace algorithms introduced in \cite{Horenko2010,Horenko2010b} (that were further developed in  \cite{gerber15,lukas18}): to get the best possible use of the algebraic structure in [\ref{Transformed}]-[\ref{c2}],  resulting algorithms should iterate between two distinct optimization problems, where the problem [\ref{Transformed}]-[\ref{c2}] is solved w.r.t. $\Lambda$ (for a current fixed values of $\Gamma$) and in the following step - w.r.t. $\Gamma$ (for current fixed values of $\Lambda$). The $\Lambda$-optimization step will involve independent solution of $\mathbf{K}$ regularized stationary Maximum Entropy problems.  As a result, an iterative procedure would converge monotonically to a local maximum solution of the problem [\ref{Transformed}]-[\ref{c2}]. This procedure - further referred to as TV-Entropy - should be repeated for different combinations of the input parameters $\mathbf{K}$, $\sigma_C$ (from some pre-defined discrete sets) - and common model discrimination tools like Akaike Information Criterion (\cite{Burnham2002}), cross-validation (\cite{Kohavi1995}) or the bootstrap (\cite{Burnham2002}, \cite{Kohavi1995}) will be deployed to identify the most optimal combination of -  hopefully small  - parameters $\mathbf{K}$, $\sigma_C$. 

It is straightforward to verify that the second term in (\ref{Transformed}) can alternatively be formulated as a set of two linear inequality constraints:
 \begin{eqnarray}
\label{c3}
&&  |\lambda^{(k)}|_1\leq C_{\lambda^{(k)}}, \indent\text{for all }  k=1,\dots,\mathbf{K}\\
\label{c4}
&&\sum_{t_1,t_2=1}^{T} \sum_{i_1,i_2=1}^{n}|\gamma^{(k)}_{t_1,i_1} - \gamma^{(k)}_{t_2,i_2}|\leq C_{\gamma}\indent\text{for all }  k=1,\dots,\mathbf{K},
\end{eqnarray}
where $C_{\gamma}\sum_{k=1}^{\mathbf{K}}C_{\lambda^{(k)}}\leq C$, with $C$ being the global TV-constant defined in (\ref{TV-norm}). This formulation allows a separate handling of constraints for the regime-switching and for the MaxEnt parameters with two separate sets of constraining variables $C_{\lambda^{(k)}}$ and $C_{\gamma}$. In contrast, the Tykhonov-like formulation (\ref{Transformed}-\ref{c2}) allows a joint simultaneous control for all of the variables by means of a single constraining variable $\sigma_C$.  

\begin{figure}[tbhp]
\centering
\includegraphics[angle=-90,width=1.1\linewidth]{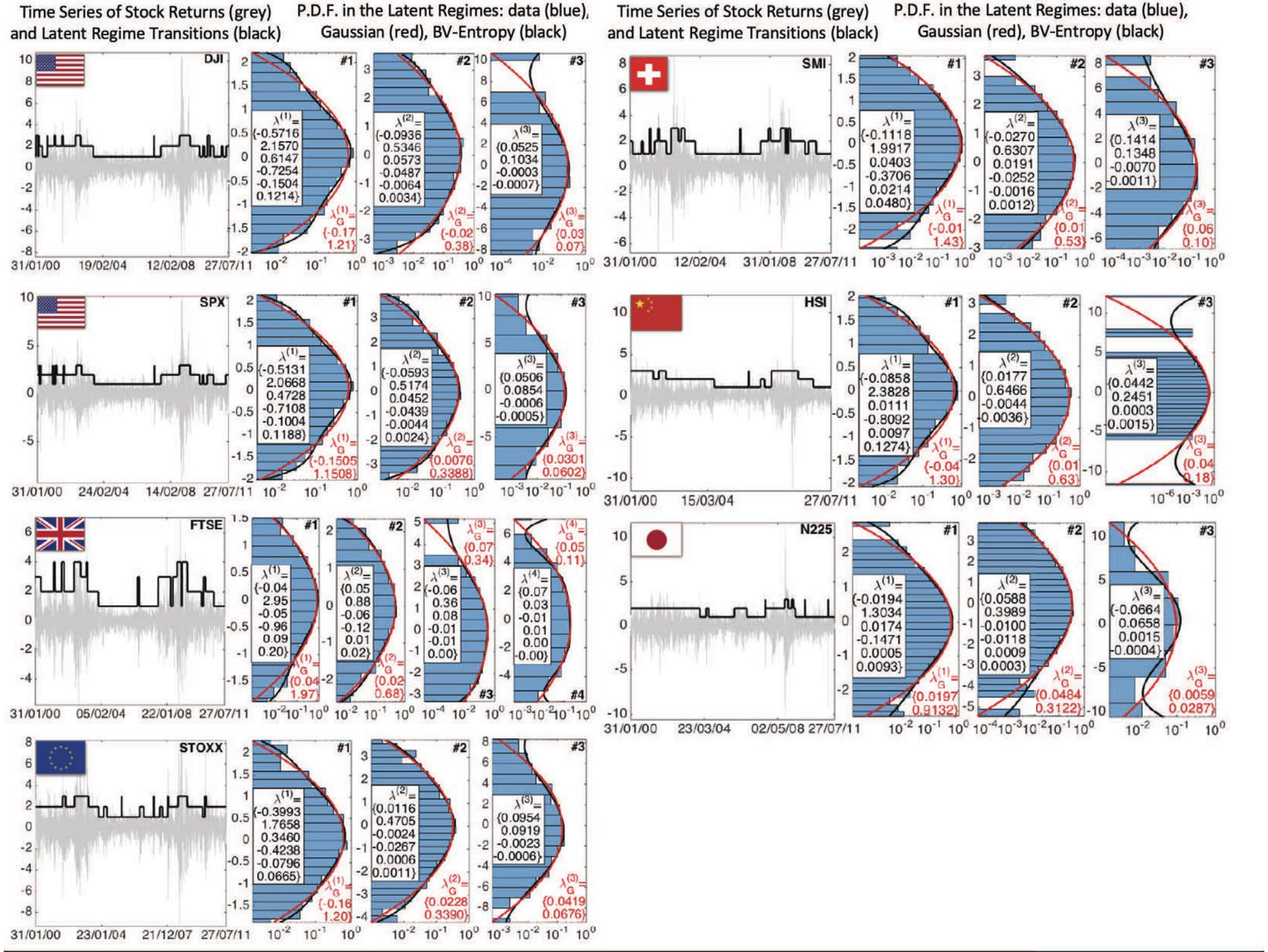}
\caption{The optimal regime-switching paths obtained by the sparse multidimensional MaxEnt-model [\ref{Transformed}]-[\ref{c2}] and the comparison of histograms of the corresponding regimes data (blue bars) to the estimated Maximum Entropy densities (black) and fitted Gaussian densities (red) for DJI, SPX, FTSE, STOXX, SMI, HSI and N225 index data.}
\label{fig:optimal_models2}
\end{figure}
\begin{figure}[tbhp]
\includegraphics[angle=-90,width=1.1\linewidth]{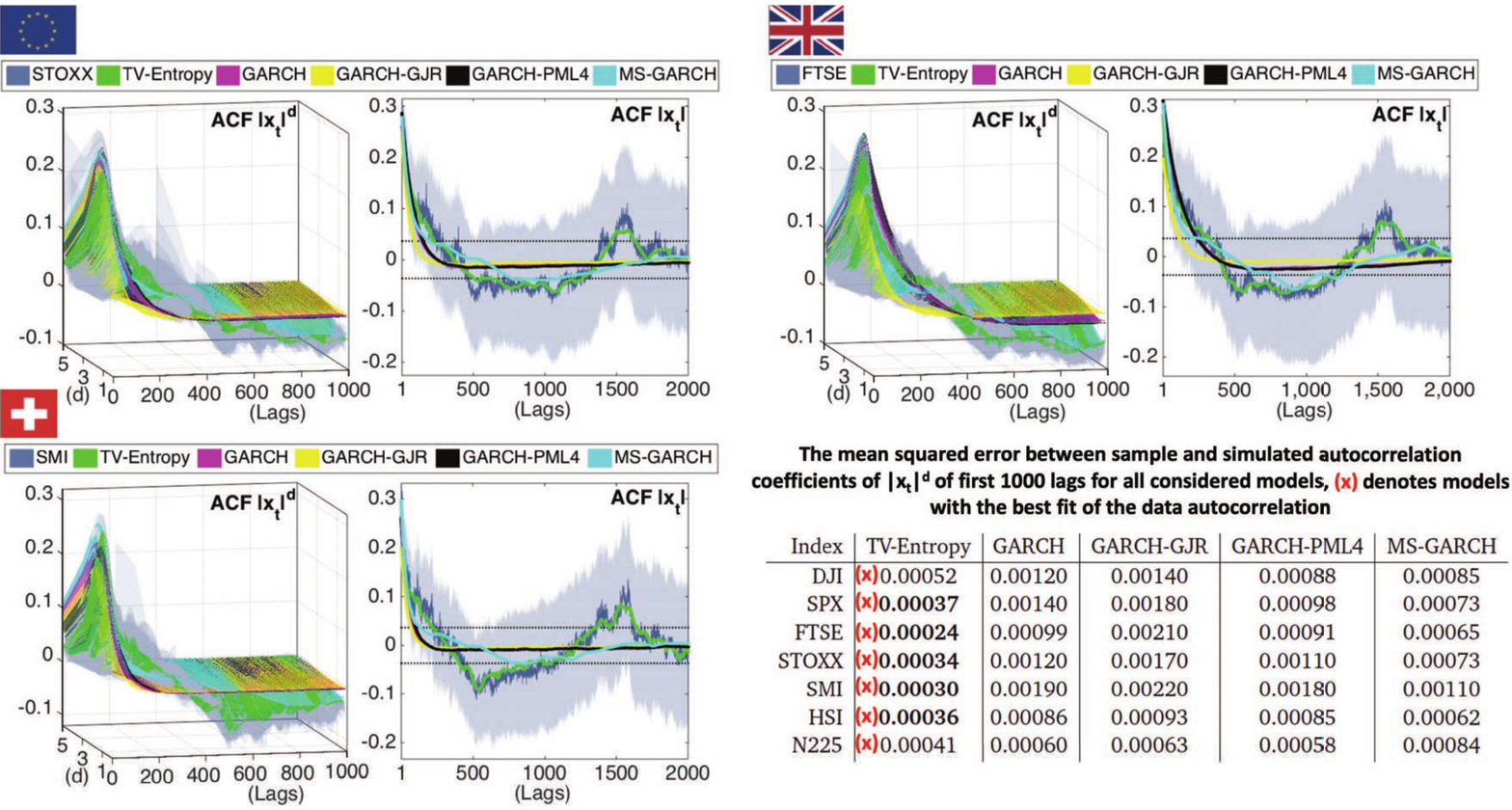}
\caption{Autocorrelation function of the absolute returns as a function of lag, power $d$ (left) and $d=1$ (right) computed from the SMI(top) and STOXX (bottom) index data and the corresponding mean values inferred from the samples generated according to the optimal parameters of all considered models. The gray dashed lines are i.i.d. confidence intervals, and the red shaded area are confidence intervals under the assumption of the MA process.}
\label{fig:ACF_DJI}
\end{figure}
\section*{Results}
First, we illustrate an application of the TV-Entropy methodology introduced above and its comparison to the common regime identification methods like HMM and the adaptive Gaussian Moving Window (GMW) in a study with artificial simulated data. For this simulated data study we use the model system proposed in \cite{marchenko2018towards}: we generate multiple samples from the regime-switching model with two Gaussian regimes, i.e.,  $\tilde{x}_t = \sum_{i=1}^2 \gamma_t^{(i)}\tilde{x}^{(i)}_t$, where $\tilde{x}^{(i)}_t \sim \mathcal{N}(0,v^{(i)})$ with $v^{(1)} = 1$, $v^{(2)} = \tilde{v}$. For each value of $v^{(2)}$ we generate 100 samples with 1000 points. The regime switching weights are discrete and satisfy the convexity constraints in [\ref{c2}], imposing only one active regime at time $t$. There are three regime transitions in data-generating process that occur every 250 points. The time-dependent variance signal can then be computed as $v_t = \sum_{i=1}^2 \gamma_t^{(i)}v^{(i)}$. This test system is designed to be favorable for HMM and GMW - since their common variants rely on the Gaussianity assumption for the realizations. The TV-Entropy models were estimated with two regimes, six density regime parameters and $[1 : 10]$ number of regime switches. We used 10 annealing steps for estimating both TV-Entropy and HMM models. The corresponding optimal parameters were chosen using BIC criteria. For GMW model we used bandwidth value $b=\{10, 30, 50\}$ to obtain various levels of persistence. As demonstrated in Figure \ref{fig:sim_study}, TV-Entropy outperforms these common models in data classification and reconstruction of the variance signal used in data-generating process. 
\begin{figure}[tbhp]
\includegraphics[width=1\linewidth]{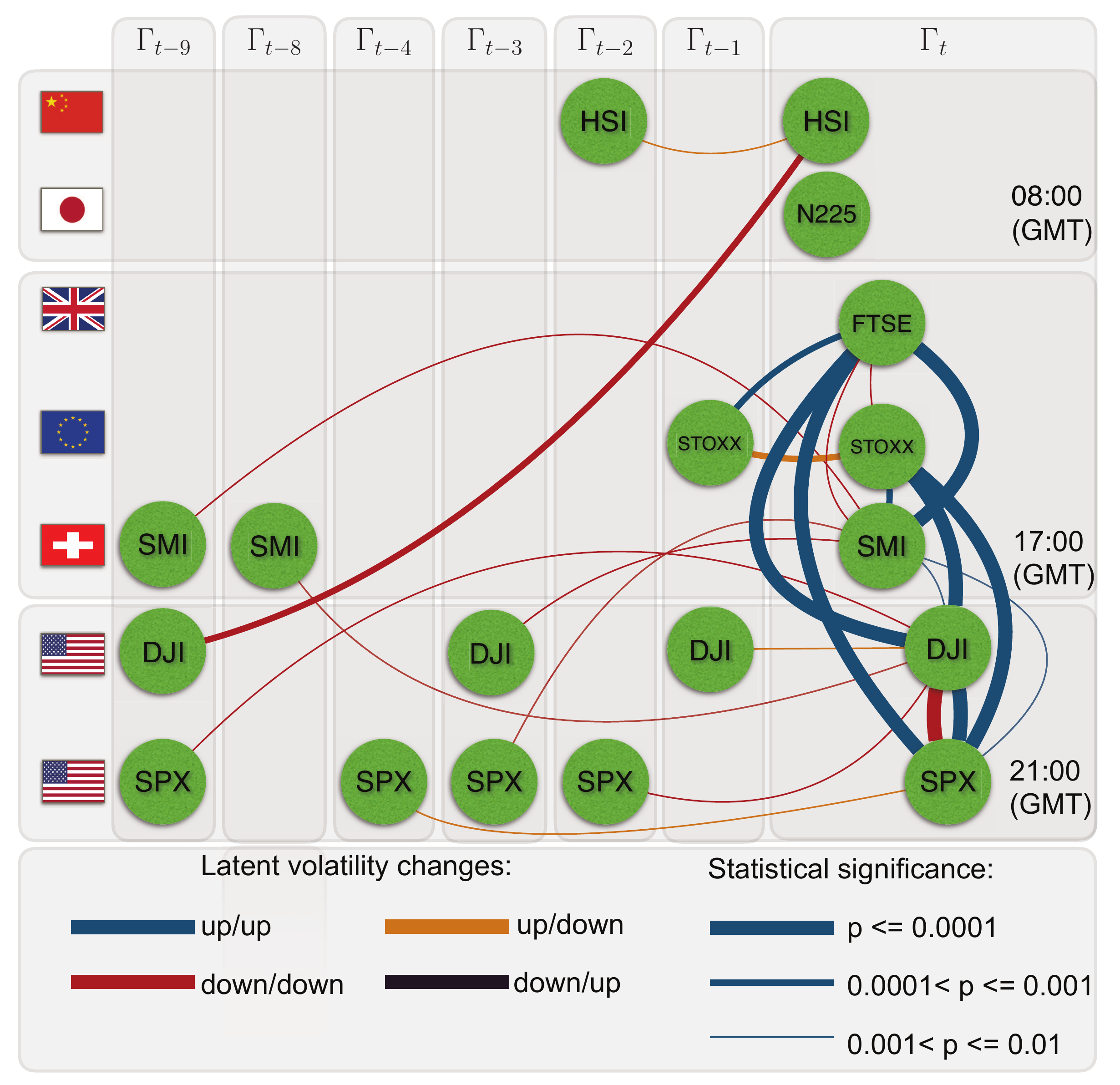}
\caption{Latent structure relation graph, showing how the regime transitions are correlated between assets over time. The connection lines are scaled according to the p-value obtained by Fisher's exact test from the vales of latent regime variables $\Gamma$ obtained with [\ref{Transformed}]-[\ref{c2}].}
\label{fig:graph}
\end{figure}
\begin{table}[h]
\centering
\caption{Classification of the volatility models used in the comparison.}
\begin{tabular}{l|r|l}\label{tab:models}
 & Stationary and without & Nonstationary and with\\
  &regime transitions &regime transitions\\\hline
 Parametric & GARCH/GARCH-GJR & MS-GARCH\\
 Semi/Nonparametric & GARCH-PML4 & TV-Entropy \\
\end{tabular}
\end{table}

In the following empirical study we compare the performance of the TV-Entropy approach  [\ref{Transformed}]-[\ref{c2}] to different popular variants of GARCH models (Table \ref{tab:models}). For the sake of simplicity, we will start with $ \sigma_C=0$. For a comparison with common econometric models we use a classic GARCH\footnote{With this abbreviation we refer to GARCH(1,1) model.} model with normally distributed innovations (\cite{Bollerslev}) and a GARCH-PML4 model that uses Maximum Entropy principle to achieve the less-restrictive description of the density (\cite{Rockinger_2011}). Additionally, we employ GARCH-GJR model that incorporates the asymmetric influence of positive and negative news on volatility (\cite{glosten1993relation}) and the MS-GARCH model that assumes the presence of several hidden GARCH regimes in data with the regime transitions governed by a Markov chain (\cite{bauwens2014marginal}). 

As benchmark problems we considered the daily percentage log-returns time series of the seven major world market indexes: DJI, SPX, FTSE, STOXX, SMI, HSI and N225\footnote{DJI (The Dow Jones Industrial Average Index, USA), SPX (The Standard \& Poor's 500 index, USA US), FTSE (The Financial Times Stock Exchange 100 Index, UK), STOXX (EURO STOXX 50 Index, EU), SMI (The Swiss Market Index, CH), HSI (The Hang Seng Index, HK/CN) and N225 (The Nikkei Stock Average Index, JP).}. The data is available in the Oxford-Man Institute's ''realised library'' \cite{gerd2009oxford}\footnote{The dataset and the source code used in numerical experiments are available at \url{https://github.com/Ganna85/TV-Entropy}}. The sample size and some related statistics of all considered samples are gathered in Table \ref{tab:samples}. All considered samples are not normally distributed \footnote{For the normal distribution the skewness and kurtosis should be equal to zero and four respectively.}. High kurtosis points to the presence of the fat tails in data and non-zero skewness suggests the asymmetry in the distribution of the returns. These properties are consistent with stylized facts often found in empirical data.

\begin{table}[tbhp]
\centering
\caption{Description of the data samples, where $T$ is a sample size}
\begin{tabular}{r|c|r|l}\label{tab:samples}
 Index & T & skewness & kurtosis \\\hline
 DJI & 2864 &  0.08 & 10.96 \\
 SPX & 2862 & -0.08 & 10.11 \\
 FTSE & 2878 & -0.09 & 6.81\\
 STOXX & 2896 & -0.12 & 7.61\\
 SMI & 2875 &  0.03 & 9.26\\
 HSI & 2603 &  0.13 & 16.38\\
 N225 & 2773 & -0.41 & 13.30\\
 \end{tabular}
\end{table}

To reduce the number of estimation routines needed to obtain the optimal parameters of the TV-Entropy, we split the estimation procedure into two stages. First, models are estimated for all combinations of the input parameters outlined in Table \ref{tab:param} without $l_1$-regularization of the regime parameters. The optimal number of the regimes ($K^*$) and transitions upper bound ($C^*_{\gamma}$) for each data set are chosen based on minimal BIC value. 

\begin{table}[tbhp]
\caption{Input parameters, where $K$ is the number of hidden regimes, $C_{\gamma}$ is the maximum number of transitions per regime, $N$ is the number of annealing steps used during estimation, $k_i$ is a number of moment constraints in the Maximum Entropy regime $i$, $C_{\lambda^{(i)}}$ is $l_1$ bound.}
\label{tab:param}
\centering
\begin{tabular}{c|c|c|c|c}
 K & $C_{\gamma}$ & N & $k_i$ & $C_{\lambda^{(i)}}$\\\hline
 \{1,2,3,4\} & \{1,2,3,$...$,50\} & 10 & 6 & +$\infty$\\ 
\end{tabular}
\end{table}

Next, in order to identify the optimal number of local parameters needed to describe data in each of the regimes, we estimate TV-Entropy models with various values of $l_1$-regularization bounds, while keeping the number of hidden regimes and transitions fixed according to their optimal values ($K^*$,$C^*_{\gamma}$) obtained at the previous step. The range of $l_1$ bounds $C_{\lambda^{(i)}}$ is data-dependent and varies across samples. To choose the appropriate range for each sample we analyze the corresponding unregularized solution. 

\begin{table}[tbhp]
\centering
\caption{Best regularized TV-Entropy models, where $K^*$, $C^*_{\gamma}$ are the optimal values of parameters with respect to the BIC, $k^*$ is the vectors with the optimal number of regime parameters}
\begin{tabular}{r|c|c|c}\label{tab:best}
 Index & $K^*$ & $C^*_{\gamma}$ & $k^*$ \\\hline
 DJI & 3 & 21 & [6,6,4]\\
 SPX & 3 & 18 & [6,6,4]\\
 FTSE & 4 & 12 & [6,6,6,6]\\
 STOXX & 3 & 16 & [6,6,4]\\
 SMI & 3 & 19 & [6,6,4]\\
 HSI & 3 & 8 & [6,4,4]\\
 N225 & 3 & 11 & [6,6,4]\\
 \end{tabular}
\end{table}

As shown in Table \ref{tab:best}, the application of the TV-Entropy model reveals three hidden regimes in all of the considered benchmarks, with an exception of the FTSE index data, where the four-regime model is identified as mBIC-optimal. The obtained regime-switching dynamic appears to be persistent for all of the considered benchmarks. Particularly, the highest number of regime transitions allowed per regime is 21 in the case of DJI and the lowest number is 8 in the case of HSI. Using $l_1$ regularizaion lead to identification of simpler models with respect to the number of regime parameters needed to describe the data. In all samples, except FTSE, we were able to eliminate irrelevant parameters, as reflected in the values of $k^*$ of Table \ref{tab:best} representing the optimal number of the parameters in each regime. The corresponding regime densities with reduced parameters (shown in Figure \ref{fig:optimal_models2}) can be interpreted as curved exponential distributions. For the interpretation of the results it is important to note that integration domain is finite and rescaled to [-1 1] interval at the time of estimation. This approach allows us to resolve integrability issues otherwise present in fitting densities with Maximum Entropy. Apart from the Maximum Entropy distributions estimated by the method (in black), we fit the Gaussian densities to the data in each regime (in red). As shown in the Table \ref{tab:samples}, the unconditional densities of all seven datasets are skewed with heavy tails. As seen from the comparison, the nonparametric TV-Entropy densities provide a better fit for fat tails and asymmetry exhibited by the underlying densities, compared to the Gaussian distribution function\footnote{The Gaussian density is fully described with two moments (e.g., the mean and the variance) and it corresponds to the Maximum Entropy distribution with first two moments constraints ($k$=2).}. This effect is most prominent in the regimes where the data points contributing to heavy mass at the tails are observed (for instance regime $\#3$ of HSI in Figure \ref{fig:optimal_models2}). 

As shown in Table \ref{tab:in_sampleLL}, the memoryless TV-Entropy model outperformed all of the finite-memory GARCH models across all seven benchmarks w.r.t. the log-likelihood and the BIC values. The highest log-likelihood value suggests the superior fit, while the lowest BIC value suggests the best balance between the fit and the complexity among all the considered models. As demonstrated by the posterior probabilities inferred from Schwartz weights [\cite{Burnham2002}], the difference in BIC values is highly significant. These results indicate that the assumption of finite autoregressive memory imposed by the GARCH models is redundant for all of the considered time series data, once nonstationary regime switches are taken into account.
    
One of the central arguments for using the GARCH models is based on their ability to fit the autocorrelation function of $|x_t|^d$ as a function of lag times and exponents $d$ (\cite{ding1993long}). As was shown in \cite{ding1993long}, there is little to no auto-correlation in daily returns. The highest auto-correlation is observed in the absolute returns and it is significant even at very large lags, suggesting a presence of autoregressive memory in the data. In the following we commence a simulation study where we draw 1000 samples from all previously obtained optimal models. We then compute the mean values of autocorrelation coefficients at each lag and compare them to the auto-correlation coefficients obtained on a real data. Specifically, we analyze the auto-correlation function of $|x_t|^d$ as a function of $d$ and a lag as shown in the left panels of Figure \ref{fig:ACF_DJI} for SMI, FTSE and STOXX index data. The summary of results for all of the  considered indexes are shown in the lower-right panel of the Figure 3. Consistently with \cite{ding1993long}, the highest serial correlation is observed around the value $d=1$ for all of the considered benchmarks. As can be seen from Figure \ref{fig:ACF_DJI}, even at the very large lags there is a significant serial correlation as the values of the coefficients lay outside the confidence intervals (light grey) constructed under the null-hypothesis that the obtained series are completely random. Results from Figure \ref{fig:ACF_DJI}  demonstrate that the simpler, conditionally memoryless TV-Entropy models (green) provide the closest approximation of the sample autocorrelation function (grey), as compared to all of the considered GARCH models. 

Next, all of the obtained models will be used to produce volatility forecasts and, consequently, the 1-day-ahead Value-at-Risk (VaR) forecasts. We compare the online VaR prediction quality for the considered methods and benchmarks. Once we observe the new data point we confirm or correct the assigned regime value $\Gamma$ using the optimal parameters obtained during the in-sample estimation. To compare the quality of the VaR forecasts, we commence the unconditional coverage test \cite{kupiec1995techniques}. As shown in Table \ref{tab:var}, there is no particular model that would be clearly preferred for all of the considered assets. However, since the results are not statistically-distinguishable (the confidence intervals are overlapping), it can be concluded that the conditionally memoryless TV-Entropy - relying on fewer tuneable model parameters - is a good alternative to the traditional GARCH models in forecasting 1-day-ahead VaR.
\begin{table}
\centering
\caption{Loglikelihood (top), BIC (middle) and posterior probabilities (bottom) values obtained for in-sample analysis}
\begin{tabular}{lrrrrr}\label{tab:in_sampleLL}
 Index & TV-Entropy & GARCH & G-GJR & G-PML4 & MS-GARCH\\
\midrule
 DJI & -3906.43 & -4058.89 & -4000.98 & -4034.36 & -4048.55\\
 SPX & -4016.21 & -4150.95 & -4091.07 & -4128.38 & -4139.24\\
 FTSE & -3465.43 & -3580.84 & -3552.85 & -3563.42 & -3577.93\\
 STOXX & -4415.74 & -4545.38 & -4473.82 & -4528.83 & -4534.32\\
 SMI & -3496.79 & -3641.14 & -3601.21 & -3625.24 & -3631.99\\
 HSI & -3443.06 & -3522.12 & -3521.07 & -3522.29 & -3508.54\\
 N225 & -4003.36 & -4100.27 & -4080.22 & -4067.87 & -4094.51\\
 \midrule
 DJI & 7948.18 & 8141.66 & 8033.80 & 8108.52 & 8113.10 \\
 SPX & 8167.73 & 8325.78 & 8213.99 & 8296.56 & 8294.47\\
 FTSE & 7129.99 & 7185.58 & 7137.57 & 7166.67 & 7171.86\\
 STOXX & 8966.99 & 9114.66 & 8979.52 & 9097.52 & 9084.64\\
 SMI & 7136.94 & 7306.16 & 7234.27 & 7290.30 & 7279.97\\
 HSI & 7004.08 & 7067.84 & 7073.59 & 7083.91 & 7047.08\\
 N225 & 8141.50 & 8224.32 & 8192.15 & 8175.38 & 8205.02\\
 \midrule
 DJI & (1.00) & (0.00) & (0.00) & (0.00) & (0.00)\\
 SPX & (1.00) & (0.00) & (0.00) & (0.00) & (0.00)\\
 FTSE & (0.98) & (0.00) & (0.02) & (0.00) & (0.00)\\
 STOXX & (1.00) & (0.00) & (0.00) & (0.00) & (0.00)\\
 SMI & (1.00) & (0.00) & (0.00) & (0.00) & (0.00)\\
 HSI & (1.00) & (0.00) & (0.00) & (0.00) & (0.00)\\
 N225 & (1.00) & (0.00) & (0.00) & (0.00) & (0.00)\\
\bottomrule
\end{tabular}
\end{table}
Next, we analyze the relationships between the latent volatility transitions inferred by TV-Entropy (see the left panel of Figure \ref{fig:optimal_models2}). Identified maximum entropy regimes are characterized by different volatility levels. As a result, the regime transitions are jumps indicating an increase or a decrease of the volatility level. For every considered asset we construct two distinct binary variables describing such a behavior, where zeros indicate no jump at a current time instance, and ones stand for transition to the regime with only increased (''up'') or only decreased (''down'') volatility. We then perform a pair-wise comparison of the obtained categorical time series (characterized by "up/up", "up/down", "down/up", "down/down" directions) using Fisher's exact test (\cite{fisher1922interpretation}) and analyze the resulting p-values to identify the statistically-significant relations between the regime transitions. Shown in Figure \ref{fig:graph} are only the most significant relationships ($p \leq 0.01$) among these categorical time series of latent regime transitions. The assets in the American and in the European regions appear to be heavily connected through their latent regime transitions. These connections are mostly governed by the volatility increase on the short time scale and volatility decrease on longer time scales. As for the considered Asian markets, the inferred latent influence of the American and European markets is present but it appears to be delayed in time and driven mostly by a decrease in the latent volatility levels. 

\section*{Discussion}
In this work we presented a sparse extension of the nonstationary MaxEnt methodology from one to multiple dimensions, aiming at identification of the most qualitative (in terms of the loglikelihood) and the least complex (in terms of the information content and the required number of tuneable parameters) representation for multidimensional time series data.

In application to analysis of financial time series we show that one of the important distinctions between the entropy-based and the common heteroschedastic approaches used in economics and finance is an assumption about the finite autoregressive memory in the underlying data-generating process. Traditionally, the observed data is assumed to be explicitly dependent on its past realizations. In the presented application of the MaxEnt-framework we do not impose additional assumptions about memory and do not include tuneable parameters that describe it. The hypothesis that realizations are independent within the regimes, has been previously explored in \cite{Bulla} and \cite{Ryden} in the context of parametric HMMs, where the authors showed that proposed models can reproduce the empirical properties of daily returns, especially in the case when conditional distributions are not normally distributed. However, also these approaches impose an explicit a priori memory assumption (Markovianity) on the level of the regime-switching process, and it remains unclear whether this assumption is necessary and/or sufficient for realistic financial data. The TV-Entropy approach reveals that the volatility for all of the considered benchmarks is best described by the time-dependent persistent process, where persistence is identified through the adaptive regularization [\ref{c2}] of the regime-switching process and from the fact that within every regime the volatility remains stationary and i.i.d.

\begin{table}
\centering
\caption{Unconditional coverage test results for 95\% and 99\% VaR. The expectation of VaR violations should be close to 5\% and 1\% respectively.}
\begin{tabular}{lrrrrr}\label{tab:var}
 Index & TV-Entropy & GARCH & G-GJR & G-PML4 & MS-GARCH\\
\midrule
 DJI & 0.044 & 0.042 & 0.043 & 0.041 & 0.043\\
 SPX & 0.045 & 0.045 & 0.047 & 0.042 & 0.046\\
 FTSE & 0.046 & 0.059 & 0.059 & 0.055 & 0.057\\
 STOXX & 0.054 & 0.052 & 0.052 & 0.048 & 0.052\\
 SMI & 0.048 & 0.049 & 0.051 & 0.047 & 0.049\\
 HSI & 0.056 & 0.054 & 0.051 & 0.054 & 0.015\\
 N225 & 0.044 & 0.047 & 0.051 & 0.044 & 0.046\\
 \midrule
 DJI & 0.013 & 0.018 & 0.015 & 0.014 & 0.018\\
 SPX & 0.013 & 0.017 & 0.018 & 0.014 & 0.017\\
 FTSE & 0.015 & 0.021 & 0.023 & 0.017 & 0.021\\
 STOXX & 0.015 & 0.017 & 0.022 & 0.014 & 0.017\\
 SMI & 0.014 & 0.018 & 0.019 & 0.014 & 0.016\\
 HSI & 0.017 & 0.020 & 0.019 & 0.020 & 0.003\\
 N225 & 0.013 & 0.016 & 0.017 & 0.012 & 0.019\\
\bottomrule
\end{tabular}
\end{table}

Starting with the intrinsically-multivariate but ill-posed MaxEnt-formulation in the formula (1-2), we derived that introducing two (mild) Assumptions 1 and 2 its solution can be approximated from below with a well-posed solution of the sparse regularized problem (10-11). Problem (10-11) is still multivariate since the hidden regime variables $\gamma_{t,i}^{(k)}$ change with the dimension $i$, time $t$ and the regime $k$. As shown in the Fig. 4, they capture the latent multivariate relation structure over different dimensions and times.

In this manuscript we analyzed seven leading world market indexes across America, Europe and Asia.
We found that nonparametric TV-Entropy approach outperforms all of the considered benchmark models for in-sample analysis in terms of the log-likelihood, simplicity (the number of free parameters), the information content (BIC and the posterior model probabilities) and in terms of quality when describing the underlying autocorreltaion function behavior. The out-of-sample study indicates that TV-Entropy methodology is an effective alternative to the  GARCH models for forecasting of the 1-day-ahead VaR. The TV-Entropy approach could closely reproduce serial correlation patterns found in data, especially at the large lags (unlike any of the GARCH models considered). This study indicates that the nonstationary MS-GARCH (combining GARCH with the Markovian regime transitions model) allows for a better description of the serial correlation than single regime GARCH models, but it is not able to match the data as accurately as the TV-Entropy model. Finally, Figure \ref{fig:graph} illustrates how the regime transition processes inferred from the data allows to identify the statistically-significant temporal relations between latent volatility level transitions across different markets.
In particular, these findings indicate that the negative news have a stronger short-term impact on the markets, as we observe that statistically most-significant latent connections are associated with the increase in volatility levels. 

\bibliographystyle{plainnat}
\bibliography{RefProj}

\begin{thebibliography}{31}
\providecommand{\natexlab}[1]{#1}
\providecommand{\url}[1]{\texttt{#1}}
\expandafter\ifx\csname urlstyle\endcsname\relax
  \providecommand{\doi}[1]{doi: #1}\else
  \providecommand{\doi}{doi: \begingroup \urlstyle{rm}\Url}\fi

\bibitem[Agmon et~al.(1979)Agmon, Alhassid, and Levine]{Agmon79}
N.~Agmon, Y.~Alhassid, and R.~D. Levine.
\newblock An algorithm for finding the distribution of {M}aximal {E}ntropy.
\newblock \emph{Journal of computational physics}, 30\penalty0 (2):\penalty0
  250--258, 1979.

\bibitem[Bauwens et~al.(2014)Bauwens, Dufays, and
  Rombouts]{bauwens2014marginal}
L.~Bauwens, A.~Dufays, and J.~V.~K. Rombouts.
\newblock Marginal {L}ikelihood for {M}arkov-switching and change-point {GARCH}
  models.
\newblock \emph{Journal of Econometrics}, 178:\penalty0 508--522, 2014.

\bibitem[Berger et~al.(1996)Berger, Della~Pietra, and
  Della~Pietra]{berger1996maximum}
A.~L Berger, V.~J. Della~Pietra, and S.~A. Della~Pietra.
\newblock A {M}aximum {E}ntropy approach to natural language processing.
\newblock \emph{Computational linguistics}, 22\penalty0 (1):\penalty0 39--71,
  1996.

\bibitem[Bollerslev(1986)]{Bollerslev}
T.~Bollerslev.
\newblock Generalized autoregressive conditional heteroskedasticity.
\newblock \emph{Journal of econometrics}, 31\penalty0 (3):\penalty0 307--327,
  1986.

\bibitem[Bulla and Bulla(2006)]{Bulla}
J.~Bulla and I.~Bulla.
\newblock Stylized facts of financial time series and hidden semi-{M}arkov
  models.
\newblock \emph{Computational Statistics \& Data Analysis}, 51\penalty0
  (4):\penalty0 2192--2209, 2006.

\bibitem[Burnham and Anderson(2003)]{Burnham2002}
K.~P. Burnham and D.~R. Anderson.
\newblock \emph{Model selection and multimodel inference: a practical
  information-theoretic approach}.
\newblock Springer Science \& Business Media, 2003.

\bibitem[Ding et~al.(1993)Ding, Granger, and Engle]{ding1993long}
Z.~Ding, C.~W.~J. Granger, and R.~F. Engle.
\newblock A long memory property of stock market returns and a new model.
\newblock \emph{Journal of empirical finance}, 1\penalty0 (1):\penalty0
  83--106, 1993.

\bibitem[Fisher(1922)]{fisher1922interpretation}
R.~A. Fisher.
\newblock On the interpretation of $\chi$ 2 from contingency tables, and the
  calculation of p.
\newblock \emph{Journal of the Royal Statistical Society}, 85\penalty0
  (1):\penalty0 87--94, 1922.

\bibitem[Gerber and Horenko(2015)]{gerber15}
S.~Gerber and I.~Horenko.
\newblock Improving clustering by imposing network information.
\newblock \emph{Science Advances (AAAS)}, 1\penalty0 (7):\penalty0 e1500163,
  2015.

\bibitem[Gerd et~al.(2009)Gerd, Lunde, Shephard, and Sheppard]{gerd2009oxford}
H.~Gerd, A.~Lunde, N.~Shephard, and K.~Sheppard.
\newblock Oxford-man institute's realized library, 2009.

\bibitem[Ggolan et~al.(1996)Ggolan, Judge, and Perloff]{ggolan1996maximum}
A.~Ggolan, G.~Judge, and J.~M. Perloff.
\newblock A {M}aximum {E}ntropy approach to recovering information from
  multinomial response data.
\newblock \emph{Journal of the American Statistical Association}, 91\penalty0
  (434):\penalty0 841--853, 1996.

\bibitem[Glosten et~al.(1993)Glosten, Jagannathan, and
  Runkle]{glosten1993relation}
L.~R. Glosten, R.~Jagannathan, and D.~E. Runkle.
\newblock On the relation between the expected value and the volatility of the
  nominal excess return on stocks.
\newblock \emph{The journal of finance}, 48\penalty0 (5):\penalty0 1779--1801,
  1993.

\bibitem[Hansen(1982)]{hansen1982large}
L.~P. Hansen.
\newblock Large sample properties of generalized method of moments estimators.
\newblock \emph{Econometrica: Journal of the Econometric Society}, pages
  1029--1054, 1982.

\bibitem[H{\"a}rdle(1990)]{Hardle}
W.~H{\"a}rdle.
\newblock \emph{Applied nonparametric regression}.
\newblock Number~19. Cambridge university press, 1990.

\bibitem[Holly et~al.(2011)Holly, Monfort, and Rockinger]{Rockinger_2011}
A.~Holly, A.~Monfort, and M.~Rockinger.
\newblock Fourth order {P}seudo {M}aximum {L}ikelihood methods.
\newblock \emph{Journal of econometrics}, 162\penalty0 (2):\penalty0 278--293,
  2011.

\bibitem[Horenko(2010{\natexlab{a}})]{Horenko2010}
I.~Horenko.
\newblock Finite element approach to clustering of multidimensional time
  series.
\newblock \emph{SIAM Journal on Scientific Computing}, 32\penalty0
  (1):\penalty0 62--83, 2010{\natexlab{a}}.

\bibitem[Horenko(2010{\natexlab{b}})]{Horenko2010b}
I.~Horenko.
\newblock On the identification of nonstationary factor models and their
  application to atmospheric data analysis.
\newblock \emph{Journal of the Atmospheric Sciences}, 67\penalty0 (5):\penalty0
  1559--1574, 2010{\natexlab{b}}.

\bibitem[Jaynes(1957)]{Jaynes1957}
E.~T. Jaynes.
\newblock Information theory and statistical mechanics.
\newblock \emph{Physical review}, 106\penalty0 (4):\penalty0 620, 1957.

\bibitem[Kohavi(1995)]{Kohavi1995}
R.~Kohavi.
\newblock A study of cross-validation and bootstrap for accuracy estimation and
  model selection.
\newblock In \emph{Ijcai}, volume~14, pages 1137--1145. Stanford, CA, 1995.

\bibitem[Kupiec(1995)]{kupiec1995techniques}
P.~Kupiec.
\newblock Techniques for verifying the accuracy of risk measurement models.
\newblock 1995.

\bibitem[Manski and McFadden(1981)]{manski1981structural}
C.~F. Manski and D.~McFadden.
\newblock \emph{Structural analysis of discrete data with econometric
  applications}.
\newblock {MIT} Press Cambridge, MA, 1981.

\bibitem[Marchenko et~al.(2018)Marchenko, Gagliardini, and
  Horenko]{marchenko2018towards}
G.~Marchenko, P.~Gagliardini, and I.~Horenko.
\newblock Towards a computationally tractable maximum entropy principle for
  nonstationary financial time series.
\newblock \emph{SIAM Journal on Financial Mathematics}, 9\penalty0
  (4):\penalty0 1249--1285, 2018.

\bibitem[Mead and Papanicolaou(1984)]{Mead1984}
L.~R. Mead and N.~Papanicolaou.
\newblock Maximum {E}ntropy in the problem of moments.
\newblock \emph{Journal of Mathematical Physics}, 25\penalty0 (8):\penalty0
  2404--2417, 1984.

\bibitem[Mora et~al.(2010)Mora, Walczak, Bialek, and Callan]{mora2010maximum}
T.~Mora, A.~M. Walczak, W.~Bialek, and C.~G. Callan.
\newblock Maximum {E}ntropy models for antibody diversity.
\newblock \emph{Proceedings of the National Academy of Sciences}, 107\penalty0
  (12):\penalty0 5405--5410, 2010.

\bibitem[Nigam et~al.(1999)Nigam, Lafferty, and McCallum]{nigam1999using}
K.~Nigam, J.~Lafferty, and A.~McCallum.
\newblock Using {M}aximum {E}ntropy for text classification.
\newblock In \emph{IJCAI-99 workshop on machine learning for information
  filtering}, volume~1, pages 61--67, 1999.

\bibitem[Phillips et~al.(2006)Phillips, Anderson, and
  Schapire]{phillips2006maximum}
S.~J. Phillips, R.~P. Anderson, and R.~E. Schapire.
\newblock Maximum {E}ntropy {M}odeling of species geographic distributions.
\newblock \emph{Ecological modelling}, 190\penalty0 (3-4):\penalty0 231--259,
  2006.

\bibitem[Posp\'{i}\v{s}il et~al.(2018)Posp\'{i}\v{s}il, Gagliardini, Sawyer,
  and Horenko]{lukas18}
L.~Posp\'{i}\v{s}il, P.~Gagliardini, W.~Sawyer, and I.~Horenko.
\newblock On a scalable nonparametric denoising of time series signals.
\newblock \emph{Communications in Applied Mathematics and Computational
  Science}, 13\penalty0 (1):\penalty0 107--138, 2018.

\bibitem[Ryd{\'e}n et~al.(1998)Ryd{\'e}n, Ter{\"a}svirta, and
  {\AA}sbrink]{Ryden}
T.~Ryd{\'e}n, T.~Ter{\"a}svirta, and S.~{\AA}sbrink.
\newblock Stylized facts of daily return series and the hidden {M}arkov model.
\newblock \emph{Journal of applied econometrics}, pages 217--244, 1998.

\bibitem[Tibshirani(1996)]{Tibshirani96}
R.~Tibshirani.
\newblock Regression shrinkage and selection via the {L}asso.
\newblock \emph{Journal of the Royal Statistical Society. Series B
  (Methodological)}, pages 267--288, 1996.

\bibitem[Wu(2003)]{Wu2003}
X.~Wu.
\newblock Calculation of {M}aximum {E}ntropy densities with application to
  income distribution.
\newblock \emph{Journal of Econometrics}, 115\penalty0 (2):\penalty0 347--354,
  2003.

\bibitem[Zellner and Highfield(1988)]{Zellner88}
A.~Zellner and R.~A. Highfield.
\newblock Calculation of {M}aximum {E}ntropy distributions and approximation of
  marginal-posterior distributions.
\newblock \emph{Journal of Econometrics}, 37\penalty0 (2):\penalty0 195--209,
  1988.

\end{thebibliography}

\end{document}